\documentclass[12pt,modern]{aastex61}
\usepackage[titletoc,toc,title]{appendix}

\shorttitle{The Main Sequence knee}
\shortauthors{Saracino et al.}

\begin{document}

\title{On the use of the main sequence knee (saddle) to measure globular cluster ages}

\author{S. Saracino} \affiliation{Dipartimento di Fisica e Astronomia,
  Universit\`a di Bologna, Via Gobetti 93/2, I-40129 Bologna, Italy}
\affiliation{INAF - Osservatorio Astronomico di Bologna, via Gobetti
  93/3, I-40129 Bologna, Italy} \correspondingauthor{Sara Saracino} \email{sara.saracino@unibo.it}

\author{E. Dalessandro} \affiliation{INAF - Osservatorio Astronomico
  di Bologna, via Gobetti 93/3, I-40129 Bologna, Italy}
\affiliation{Dipartimento di Fisica e Astronomia, Universit\`a di
  Bologna, Via Gobetti 93/2, I-40129 Bologna, Italy}

\author{F. R. Ferraro} \affiliation{Dipartimento di Fisica e
  Astronomia, Universit\`a di Bologna, Via Gobetti 93/2, I-40129
  Bologna, Italy}

\author{B. Lanzoni} \affiliation{Dipartimento di Fisica e Astronomia,
  Universit\`a di Bologna, Via Gobetti 93/2, I-40129 Bologna, Italy}

\author{L. Origlia} \affiliation{INAF - Osservatorio Astronomico di
  Bologna, via Gobetti 93/3, I-40129 Bologna, Italy}

\author{M. Salaris} \affiliation{Astrophysics Research Institute,
  Liverpool John Moores University, 146 Brownlow Hill, Liverpool L3
  5RF, UK}

\author{A. Pietrinferni} \affiliation{INAF - Osservatorio Astronomico di 
	Teramo, via M. Maggini, 64100 Teramo, Italy} 

\author{D. Geisler} \affiliation{Departamento de Astronom\'ia,
  Universidad de Concepci\'on, Casilla 160-C, Concepci\'on, Chile}

\author{J. S. Kalirai} \affiliation{Space Telescope Science Institute,
  3700 San Martin Drive, Baltimore, MD 21218, USA} \affiliation{Center
  for Astrophysical Science, John Hopkins University, Baltimore, MD
  21218, USA}

\author{M. Correnti} \affiliation{Space Telescope Science Institute,
  3700 San Martin Drive, Baltimore, MD 21218, USA}

\author{R. E. Cohen} \affiliation{Space Telescope Science Institute,
  3700 San Martin Drive, Baltimore, MD 21218, USA}
 
\author{F. Mauro} \affiliation{Instituto de Astronom\'ia, Universidad
  Cat\'olica del Norte, Av. Angamos 0610, Antofagasta, Chile}

\author{S. Villanova} \affiliation{Departamento de Astronom\'ia,
  Universidad de Concepci\'on, Casilla 160-C, Concepci\'on, Chile}

\author{C. Moni Bidin} \affiliation{Instituto de Astronom\'ia,
  Universidad Cat\'olica del Norte, Av. Angamos 0610, Antofagasta,
  Chile}

\begin{abstract} 
In this paper we review the operational definition of the so-called
main sequence knee (MS-knee), a feature in the color-magnitude diagram (CMD) 
occurring at the low-mass end of the MS. The magnitude of this feature
is predicted to be independent of age at fixed chemical
composition. For this reason, its difference in magnitude with respect to the MS turn-off
(MS-TO) point has been suggested as a possible diagnostic to estimate absolute globular cluster (GC) ages. 

We first demonstrate that the operational definition of the
MS-knee currently adopted in the literature refers to the
{\it inflection point} of the MS (that we here more appropriately
named MS-saddle), a feature that is well distinct from the knee and
that cannot be used as its proxy. 
The MS-knee is only visible in near-infrared CMDs, while the MS-saddle can be also detected in optical-NIR CMDs.

By using different sets of
isochrones we then demonstrate that the absolute magnitude of the MS-knee varies
by a few tenths of a dex from one model to another, thus showing that at the moment stellar models 
may not capture the full systematic error in the method.

We also demonstrate that while the absolute magnitude of the MS-saddle is almost coincident in different models,
it has a systematic dependence on the adopted color combinations which is not predicted by stellar models. 

Hence, it cannot be used as a reliable reference for absolute age determination.
Moreover, when statistical and systematic uncertainties are properly taken into account, 
the difference in magnitude between the MS-TO and the MS-saddle
does not provide absolute ages with better accuracy than other methods like the MS-fitting. 
\end{abstract}
\date{May 3, 2018}


\section{Introduction}
Globular clusters (GCs) are among the oldest stellar aggregates in the
Universe and are therefore 
pristine fossils of the very early epoch of galaxy formation. A detailed study of
their ages is useful for several reasons: {\it relative} ages help us to
understand the formation and the assembly chronology of the different
components (halo, disk, bulge) of our Galaxy (\citealp{Ros99};
\citealp{Zoc03}, \citealp{Mar09}); {\it absolute} ages set a lower limit to the age of
the Universe (\citealp{Buo98}, \citealp{Ste99}, \citet{Gra03}) and
provide robust constraints to the physics adopted in stellar
evolutionary models (\citealp{SW98}; \citealp{Cas99}; \citealp{Van08};
\citealp{Dot08}). Several methods have been used so far to estimate
the relative and absolute ages of these systems, mainly based on the
analysis of their optical CMDs. 

Relative
ages can be derived by using ``differential'' parameters, built from the
magnitude or the color of the main sequence turn-off point (MS-TO,
which systematically varies with time) and the magnitude or the color
of a ``reference''
feature in the CMD that is independent of the cluster
age. Examples are: (1) the {\it horizontal} parameter, defined as the
difference in color between the MS-TO and the red giant branch (RGB) at
2.5 magnitudes above the MS-TO level (see \citealp{Sar90,sd90,van90}); (2)
the {\it vertical} parameter, based on the difference in magnitude
between the horizontal branch (HB), typically measured at the RR Lyrae
instability strip, and the MS-TO (\citealp{IR84}; \citealp{Buo98}; \citealp{Ros99};
\citealp{Ste99}; \citealp{DeA05}). Of course
differential parameters have the advantage of being independent of
distance and reddening. 

Cluster absolute ages are generally estimated
by measuring the luminosity of the MS-TO point in the CMD, or by
applying the isochrone fitting method. The latter has been recently used by
\citet{Sara16} and \citet{Cor16}, to determine sub-Gyr absolute ages 
using a chi-squared minimization and a maximum-likelihood technique, respectively.
Absolute ages can also be derived from the direct comparison
of the differential parameters with the corresponding theoretical
predictions. In these cases, various sets of theoretical models (e.g., \citet{Pie04};
\citealp{SW02}; \citealp{Dot10}; \citealp{Van13})
need to be considered in order to investigate the effects of different
assumptions in the models. This effect becomes quite important when low-mass stars 
are considered in the analysis. More in general, the reliability of the absolute
 ages derived from classical methods almost totally depends on the accuracy 
 of the adopted distances and chemical abundances.

In the recent years, advanced instruments and techniques, such as
adaptive optics systems
mounted on 8-10 m class telescopes, and high resolution cameras on
board the Hubble Space Telescope (HST), 
make also near-infrared (NIR) observations very promising in the estimate of GC ages. In fact,
deep NIR photometry revealed the existence of a well-defined knee at
the lower end of the MS, approximately three magnitudes below the
MS-TO (see, e.g., the cases of $\omega$ Centauri, \citealp{Pul98}; M4,
\citealp{Pul98}; \citealp{Mil14}; NGC 3201, \citealp{Bon10}; NGC 2808,
\citealp{Mil12}; \citealp{Mas16}; 47 Tucanae, \citealp{Kal12}; M71, \citealp{DiC15}; M15,
\citealp{Mon15}). 

The MS-knee 
arises from a redistribution of the
emerging stellar flux due to an opacity change, mainly caused by the
collision-induced absorptions of molecular hydrogen in the surface of
cool dwarfs (\citealp{Lin69}; \citealp{Sau94} and references therein),
which moves low-mass MS stars towards bluer colors.\footnote{Figure 16 of \citet{CV14} shows 
that the shape of the low-mass end of the MS sensibly depends on the $\alpha$-element abundance,
suggesting that for $[\alpha/Fe]=-0.4$ a MS-knee is not present. However, 
in the $\alpha$-element abundance regime of Galactic GCs, a MS bending 
can be identified at all metallicities.} Since its
magnitude is predicted to be independent of cluster age at fixed
chemical composition, the MS-knee provides a potential anchor
(alternative to the HB luminosity) at which the MS-TO magnitude can be
referred to define a new {\it vertical} method for the age
determination, independent of GC distance and reddening. With
respect to the traditional {\it vertical} method referred to the HB
level, it has the drawback of requiring the detection of a much
fainter feature in the CMD (fainter by $\sim 7$ magnitudes), but {\it i)} it is much more
 populated because the stellar luminosity function increases towards lower masses and {\it ii)} it
should be less affected by model uncertainties, such as the treatment of
convection \citep{SM08}. For these reasons \citet{Bon10} proposed a new parameter
(hereafter $\Delta^{\rm knee}_{\rm TO}$) to measure relative and
absolute GC ages from the magnitude difference between the MS-TO and
the MS-knee levels. The method has been already tested on a few
clusters (as NGC 2808, \citealp{Mas16}; M71, \citealp{DiC15} and M15,
\citealp{Mon15}), with the conclusion that it can provide absolute age estimates 
at sub-Gyr accuracy, a factor of two better than what can be obtained with
classical methods. Relative age studies using the $\Delta^{\rm knee}_{\rm TO}$ 
method have not been performed yet.

This paper provides an in-depth analysis of both the operational
definition of the MS-knee and the potential use of this feature to
estimate absolute GC ages. 
Our study is based on both widely used stellar models and photometric observations 
of the low-mass MS in two well known GCs.

In Section 2 we summarize the diagnostic tools used to perform the analysis.
In Section 3 we review the operational definition of the MS-knee adopted in 
the literature, suggesting a more appropriate nomenclature: we name MS-knee 
the point in a NIR CMD where the MS bends to the blue, while we name MS-saddle 
the point where the MS changes curvature. 
In Section 4 we describe the procedure followed in order to measure the MS-saddle in 
the two GCs 47 Tucanae and NGC 6624, and in Section 5 we discuss their 
derived ages and uncertainties. In Section 6 we draw our conclusion.

\section{Diagnostic tools}       

\begin{figure}
\epsscale{1.}
\plotone{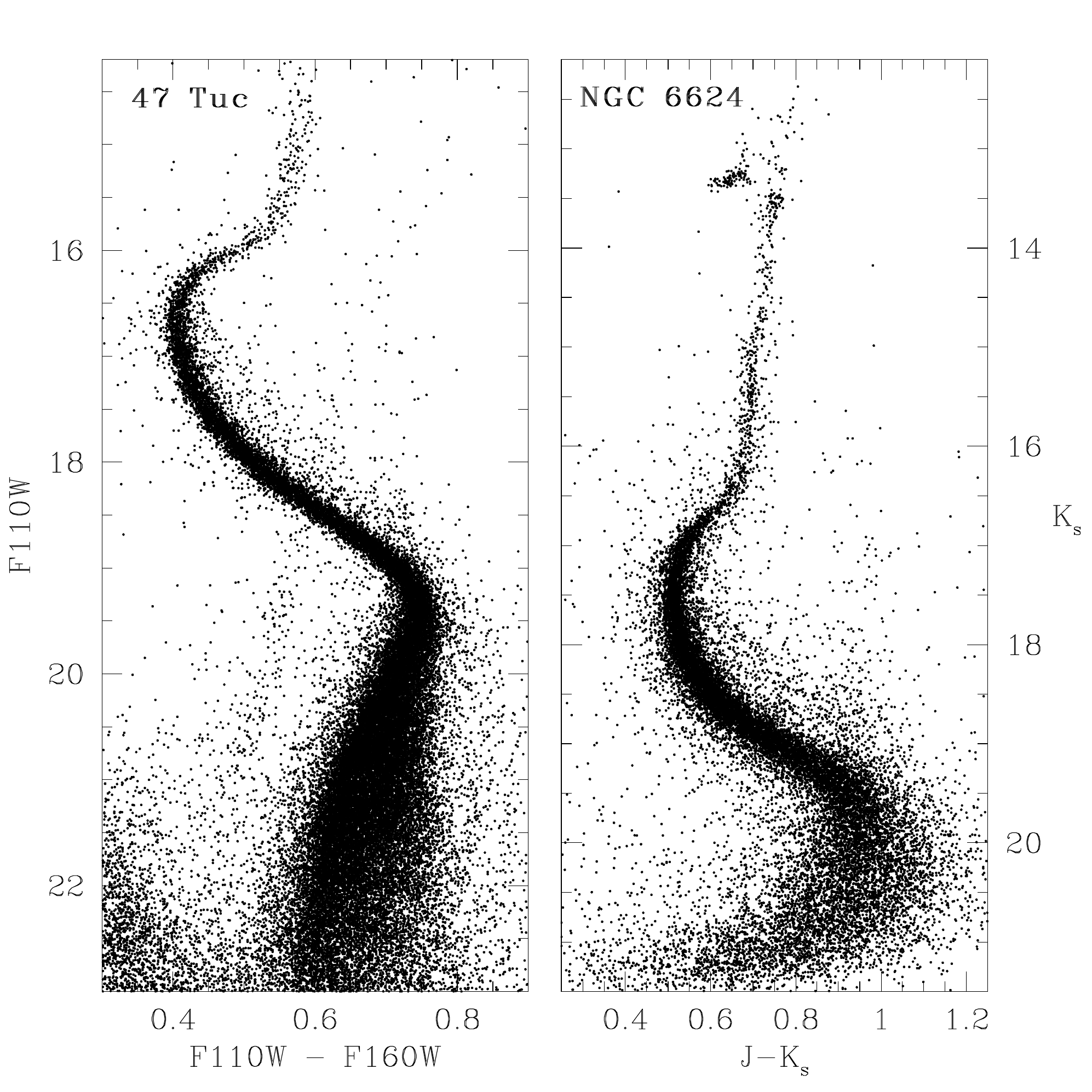}
\caption{ {\it Left panel --} ($F110W, F110W-F160W$) CMD of 47 Tucanae
  from HST observations. {\it
    Right panel --} ($K_{s}, J-K_{s}$) CMD of NGC 6624 from deep
  observations acquired with a ground-based adaptive optics
  system.}
\label{fig1}
\end{figure}

Our analysis of the low-mass MS and its most important features has been performed 
by using the following theoretical and observational tools:
\begin{itemize}
\item {\it Stellar models} - 
We considered three different sets of $\alpha$-enhanced stellar models, namely A Bag of Stellar Tracks and Isochrones (BaSTI; 
\citealp{Pie04}), the Dartmouth Stellar Evolutionary Database (DSED; \citealp{Dot07}) 
and the Victoria-Regina isochrones (VR; \citealp{Van14}). 
The BaSTI NIR colors and magnitudes are on the Johnson-Cousins-Glass photometric system, 
while the DSED and VR isochrones are on the 2MASS photometric system.
Hence, for a coherent comparison, we adopted the 2MASS photometric system as reference and
we converted the BaSTI NIR colors first on the \citet{BB88} system and then, 
by using the transformations of \citet{Car01}, into the 2MASS photometric system \citep{Cut03}.\\ 
BaSTI, DSED and VR isochrones assume different solar mixture abundances, which are based on \citet{grev93}, \citet{grev98} and \citet{Asp09}, respectively.
To obtain the same chemical content, in terms of [M/H], for the three sets of isochrones, we adopt the same $\alpha$-element abundance ([$\alpha$/Fe] = +0.4) but we are forced to adopt slightly different [Fe/H] (see Table 1 for a detailed comparison of the adopted isochrones).
We note that this is an approximation as other metals (e.g. differences in the C+N+O abundance) can play a role. However, we stress that small changes to the assumed metal abundances are expected to have much smaller effects on isochrones along the MS than other factors, such as the color-temperature transformations to the observed diagrams (see below).

\item {\it Observed GCs} - 
Deep and accurate photometry of the low-mass MS of two GCs, namely 
47 Tucanae and NGC 6624, having similar chemical composition ([Fe/H]=-0.69 and -0.60 respectively; 
see \citet{Cor16} and reference therein and \citet{Sara16} and reference therein), has been used as an empirical test bench 
for our analysis.
For 47 Tucanae we used the catalog presented in \citet{Kal12}, based on 
images acquired using the infrared channel of the Wide Field Camera 3 (WFC3)
on board the HST in the $F110W$ and $F160W$ filters (GO-11677 PI: Richer).
The resulting ($F110W, F110W-F160W$) CMD is shown in the left panel of Figure \ref{fig1}. 
It exhibits very well defined evolutionary sequences, from the base
of the RGB to the low-mass end of the MS, reaching $\sim$ 3 magnitudes
below the MS-knee. The photometric precision is of the order of a few
thousandths of a magnitude over the luminosity range sampled: $\approx$
0.002 mag at the MS-TO level, $\approx$ 0.005 at the MS-knee.
Since the \citet{Kal12} catalog samples only the external regions of
the cluster, no optical data of comparable depth are available.
In the case of NGC 6624, we have used the catalog described in
\citet{Sara16}. It has been obtained by using $J$ and $K_{s}$ images of the central regions of the cluster
acquired with the multi-conjugate adaptive optics system GeMS at the
Gemini South Telescope in Chile, as part of the proposal GS-2013-Q-23 (PI:
D. Geisler).  A detailed description of the observations and the
data-reduction procedure is reported in Section 2 of \citet{Sara16}. 
The ($K_{s}, J-K_{s}$) CMD of NGC 6624 is presented in the
right panel of Figure \ref{fig1}. 
It spans a range of more than 8 magnitudes, from the HB level down to $K_{s} \approx 21.5$. In this case,
the photometric errors are of $\approx$ 0.005 mag at the MS-TO level and $\approx$ 0.035
mag at the MS-knee. 
We have also combined the GEMINI catalog
of \citet{Sara16} with the optical HST-Advanced Camera for Survey
(ACS) catalog of \citet{Sar07}, providing $V$ and $I$ data for the
stars in common. 
\item {\it CMDs} - We investigated the low-mass MS properties 
in various CMDs, namely the ($F110W, F110W-F160W$) CMD, where theoretical isochrones are compared with HST NIR photometry of 47 Tucanae,
and in the ($K_{s}, J-K_{s}$), ($K_{s}, I-K_{s}$), and ($K_{s}, V-K_{s}$) CMDs, where theoretical isochrones are 
compared with ultra-deep NIR (ground-based) and optical (HST) photometry of NGC 6624.      
Note that the hybrid optical-NIR CMDs are the most used in the literature to detect the MS-saddle 
(see \citealp{Bon10}, \citealp{Mon15}, \citealp{DiC15} and \citealp{Mas16}).
\end{itemize} 

\begin{figure}
	\epsscale{1.}
	\plotone{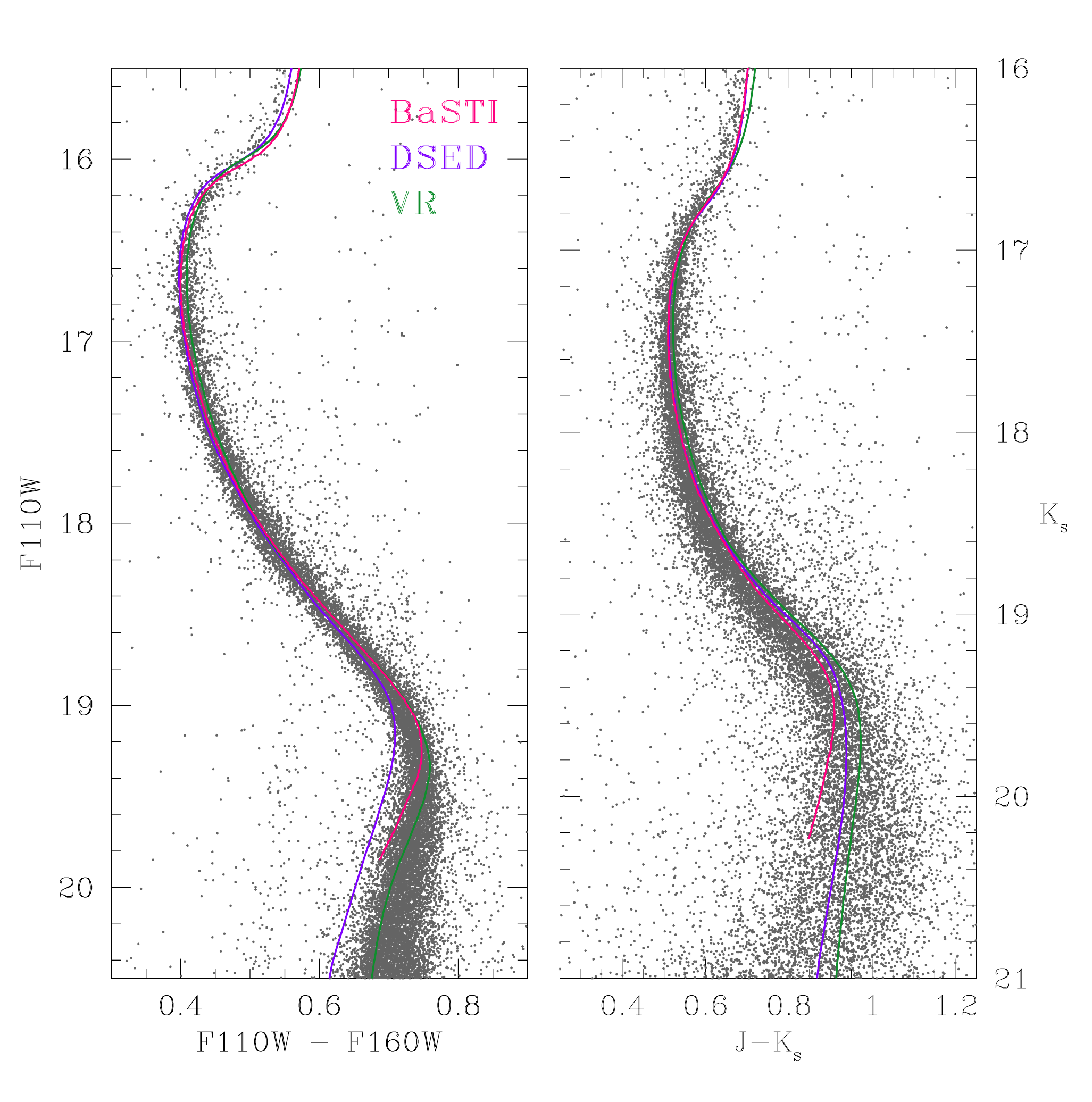}
	\caption{{\it Left panel --} ($F110W, F110W-F160W$) CMD of 47 Tucanae, with BaSTI, DSED and VR isochrones superimposed in magenta, violet and green respectively. {\it
			Right panel --} ($K_{s}, J-K_{s}$) CMD of NGC 6624 with the adopted isochrones overimposed. The color code is shown in the legend.}
	\label{fig12}
\end{figure}

In Figure \ref{fig12} we show a comparison of the ($F110W, F110W-F160W$) CMD for 47 Tucanae and of the ($K_{s}, J-K_{s}$) CMD for NGC 6624 with theoretical models. We adopted the following parameter values from the literature: $t = 11.5$ Gyr, $E(B-V) = 0.04$, $(m-M)_{0} = 13.31$ for 47 Tucanae \citep{Cor16} and $t = 12.0$ Gyr, $E(B-V) = 0.28$, $(m-M)_{0} = 14.49$ for NGC 6624 \citep{Sara16}. The color code is the same in both panels, where the magenta, violet and green lines refer to the adopted BaSTI, DSED and VR isochrones, respectively. As can be seen, the general agreement in MS between our data and the theoretical models is quite satisfactory\footnote{We note a mismatch between theoretical models and observations for magnitudes brighter than the sub-giant branch/RGB base. As already discussed in \citet{Sara16}, this is a well-known problem which might be related to some issues in the color-temperature transformations (see \citealp{Sal07}; \citealp{Bra10}; \citealp{Coh15}).}. Few exceptions: in the ($F110W, F110W-F160W$) CMD the DSED isochrone turns out to be bluer than the other two models at magnitudes fainter than $F110W = 19$, while in the ($K_{s}, J-K_{s}$) CMD a similar behavior is observed for the BaSTI isochrone at $K_{s} > 19.5$, but with a smaller discrepancy.

\section{A knee or a saddle?}

\begin{figure}
\epsscale{1.}
\plotone{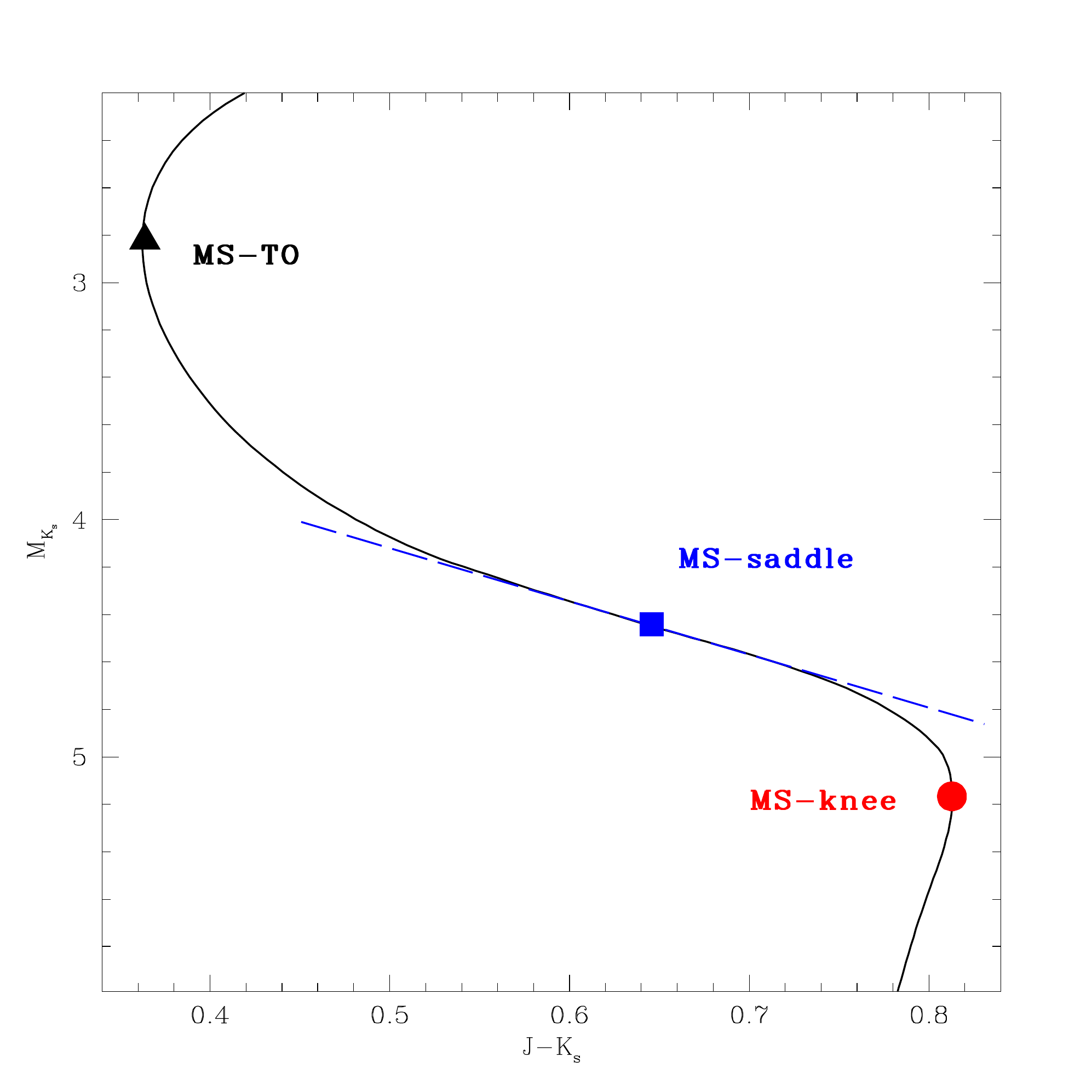}
\caption{Location of the MS-TO (black triangle), MS-knee (red circle)
  and MS-saddle (blue square) along a 12 Gyr old isochrone extracted
  from the family of \citet{Van14}. The MS-knee is here defined as the
  reddest point along the MS MRL. The MS-saddle is the point where the
  MS MRL changes shape, from convex to concave, and thus shows the
  minimum curvature. A dashed line tangential to the isochrone is
  shown at the MS-saddle point to better illustrate the morphological
  meaning of this point.}
\label{fig2}
\end{figure}

As shown in Figure \ref{fig2}, in a pure NIR ($K_{s}, J-K_{s}$) filter combination, the lowest portion
of the MS bends to the blue and creates a well defined
knee. Consistently, the MS-knee (marked with a red circle
in the figure) corresponds to the reddest point of
the MS at magnitudes fainter than the MS-TO. However, the MS-knee has been originally
defined as the point of minimum curvature along the low-mass end of
the MS ridge line (see \citealp{Bon10}). This is an inflection point
(that we name {\it MS-saddle}), where the MS ridge line changes
its curvature from convex to concave.  While it is related to 
the presence of the knee, it is certainly not coincident with it. This
is clearly illustrated in Figure \ref{fig2}, where the location of the two
features is marked along a 12 Gyr old VR isochrone: the
MS-saddle (blue square, characterized by $\approx 0.65\; M_{\odot}$ stellar mass) 
is more than 0.7 mag brighter than the MS-knee (red circle, characterized by $\approx 0.55\; M_{\odot}$ stellar mass).

However, the MS-saddle can be of interest.
In fact, as shown in Figure \ref{fig3}, while the MS-knee only occurs in NIR CMD (red
circle in the left panel) and it is not definable in hybrid optical-NIR CMDs
(middle and right panels), the MS-saddle (blue squares) can be measured
in all the diagrams. Hence, a detailed comparison between the two
features is worth investigating, in particular to assess whether the MS-saddle 
can be considered as a proxy of the MS-knee and it can be used for
measuring cluster ages.

To this end, in Figure \ref{fig4} we compare the BaSTI, DSED and VR isochrones at a fixed 
age of $t=12$ Gyr in the ($K_{s}, J-K_{s}$) CMD (left panel) and in the ($F110W, F110W-F160W$) CMD (right panel).
Triangles, squares and circles mark the MS-TO, the MS-saddle and the MS-knee, respectively, along each isochrone. 
As can be seen, the three models predict the same location in the CMD for the MS-TO and the MS-saddle, but
significantly different positions of the MS-knee. As a consequence, for the same age and metallicity, the three isochrones predict values of $\Delta ^{\rm knee}_{\rm TO}$ differing by 0.26 magnitudes in the $K_s$ filter, ranging from 2.08 to 2.36. 
A similar difference has been also measured in the $F110W$ filter, indicating that such a discrepancy 
among different models does not depend on the used NIR filters and filter combinations.

In Figure \ref{fig5} (left panel), the color and the magnitude of the MS-knee points shown in Figure \ref{fig4} are translated in effective temperature ($T_{eff}$) and luminosity ($L$), respectively. As can be seen, they are located in a region where stellar models differ in shape also in the Hertzsprung - Russell (HR) diagram.\\
To quantify such differences, we selected a reference model (BaSTI) and we measured the difference in effective temperature ($\Delta T_{eff}$) at fixed luminosity between the two model pairs (BaSTI - DSED) and (BaSTI - VR), respectively. The results are shown in the right panel of Figure \ref{fig5}, where differences up to $\Delta T_{eff}$ $\approx$ 100 K are observed both at the MS-TO and MS-knee levels.\\
The $\Delta T_{eff}$ at the MS-knee is mostly due to the uncertain choice of the stellar model boundary conditions, as discussed e.g. by \cite{chen2014}. One can clearly see in their Figures 5, 6, 9 and 10, that a small change of the boundary conditions affects the position of theoretical GCs isochrones in the ($L-T_{eff}$) diagram at the typical bolometric luminosities of the MS-knee. The difference observed at the MS-TO can be attributed to differences in the absolute C+N+O abundance and adopted physics among the three models.\\
To quantify the impact that a variation in temperature has on the position of the MS-TO and the MS-knee points, we used a VR isochrone as reference. We applied shits in temperature of $\pm$ 100 K and then we transformed it into the observed diagram by using the \citet{CV14} transformations. The results are presented in the left panel of Figure \ref{fig6}, in the ($F110W, F110W-F160W$) filter combination.
A $\Delta T_{eff}$ = $\pm$ 100 K translates into a difference of $\pm$ 0.02 mag in the MS-TO position and a difference of $\pm$ 0.15 mag in the MS-knee position. However, these values are able to explain only half of the observed discrepancies at both levels. \\
In this context, it is worth to note that BaSTI, DSED and VR isochrones are based on different model atmospheres (BaSTI - \citet{CK94}; DSED - PHOENIX \citep{hus13} and VR - MARCS \citep{G08}), thus they use different bolometric corrections (BCs) and color-temperature transformations (see e.g. \citealp{Sal07}; \citealp{Bra10}; \citealp{Coh15}).\\
In Figure \ref{fig6}, right panel, we show the effect of adopting different BCs. In particular, we compare a BaSTI isochrone with the corresponding one after applying BCs from \citet{CV14}. As can be seen, also BCs play a role in shaping the low-mass MS at NIR wavelengths. In fact, they account for differences of about 0.1 mag, 0.02 mag and 0.04 mag at the MS-knee, MS-saddle and MS-TO, respectively, thus becoming an important source of uncertainty. \\
The interplay between variations in $T_{eff}$ among models and adopted BCs produce final discrepancies of more than 0.2 mag on the MS-knee position, of about 0.05 mag and 0.04 mag on the MS-saddle and MS-TO positions, respectively, which are fully consistent with what observed both in Figures \ref{fig4} and \ref{fig5}. 

Major progress in model atmosphere calculations for low-mass and very
low-mass stellar models (that provide
both BCs and model boundary conditions) is needed for
firmer theoretical predictions about the MS-knee.

Hence, at the moment, because of the large uncertainties
currently affecting the theoretical models, the MS-knee can be safely adopted only for relative age studies 
where a comparison between different stellar models is not necessary. 

The location of the MS-saddle appears to be
much more stable (with magnitude variations among different models smaller than 0.05 mag in $K_{s}$ and 
in the corresponding bolometric luminosity, see Figures \ref{fig4} and \ref{fig5}), demonstrating that the MS-saddle is not very sensitive to the
morphology and the location of the knee. Hence the first result of
this investigation is that the MS-saddle cannot be used as a proxy for
the MS-knee. However its reduced model-dependence 
and its potential measurability also in combined
optical-NIR CMDs call for a thorough investigation of its reliability
and stability as a reference feature to estimate cluster ages.  

Of course, the first requisite is that, for fixed chemical composition,
the magnitude of the MS-saddle is independent of the cluster age. This is
indeed confirmed by all the stellar models considered above. The
predicted $K_{s}$-band magnitude of the MS-saddle (and, for comparison, of
the MS-knee) for stellar populations with ages ranging from 9.5 Gyr to
13.5 Gyr is shown in Figure \ref{fig7} for the three families of adopted
stellar models. The constancy of the MS-saddle magnitude with varying the 
isochrone age indicates that, in principle, it can be used as an
anchor to which the MS-TO can be referred and used to measure both relative and absolute 
ages\footnote{The bottom panel of Figure \ref{fig7} shows that also the MS-knee
magnitude is constant for varying cluster ages. However its
value (and thus the parameter $\Delta^{\rm knee}_{\rm TO}$)
significantly depends on the adopted family of isochrones, thus
making unacceptably model-dependent any age estimate.}. 
However, since it is a geometric point, dependent on the morphology of the MS ridge line,
it can be a somewhat fragile feature both from an
observational and a theoretical point of view, requiring detailed investigation.\\ 

\section{Measuring the MS-saddle}

The first step to locate the MS-saddle in an observed CMD of a GC is to determine the
cluster mean-ridge line (MRL). 

The catalog adopted for 47 Tucanae was 
already cleaned from spurious objects \citep{Kal12}. 
In the case of NGC 6624 CMDs, a selection in the stellar sharpness parameter
(defined as in \citealp{Ste89}) was applied. We
divided the sample of stars in our catalog in 0.5 magnitude-wide bins
and for each bin we computed the median sharpness value and its
standard deviation ($\sigma$). Only stars with sharpness parameter
lying within 6$\sigma$ from the median were flagged as
``well-measured''.

The ``clean'' photometric catalogs were used to determine the MRL in
each of the four considered CMDs, namely ($F110W, F110W-F160W$) for 47 Tucanae and 
($K_{s}, J-K_{s}$), ($K_{s}, I-K_{s}$) and ($K_{s}, V-K_{s}$) for NGC 6624. 

This was done by using three different methods in order to evaluate the impact of slightly
different MRLs on the location of the MS-saddle and to derive the
precision achievable in measuring this feature.

{\it Method 1: Static bins -} we considered different magnitude bins (in
$F110W$ for 47 Tucanae and $K_{s}$ for NGC 6624) and we computed the mean
color\footnote{In order to be less sensitive to outliers (e.g. binaries, field stars), the median color 
was also derived. Negligible differences have been found with respect to the mean color due to the very well 
cleaned catalogs.} of all the stars falling in each bin, by applying an iterative
2$\sigma$-rejection procedure. We allowed the bin size to vary from
0.10 mag (lower limit, necessary to have a reasonable number of stars
per bin) to 0.50 mag (upper limit, imposed to keep an accurate
sampling of the fiducial line), in steps of 0.01 mag. At the end of
the procedure, for each of the four filter combinations, we had 41
(differently sampled) MRLs, which have been re-sampled with a 0.01
mag-stepped cubic spline.

{\it Method 2: Dynamic bins -} This method uses bins of constant size in magnitudes partially overlapping.
This means that,
at any fixed bin size, the MRLs derived from dynamic bins are more
densely sampled (at higher resolution) than those obtained from static
bins. In this case we modified the bin size from 0.10 mag to 0.50 mag
in steps of 0.05 mag, obtaining a sampling of 0.05 mag for each MRL. The resulting
9 MRLs per each filter combination have also been re-sampled with a cubic
spline of 0.01 mag steps.

{\it Method 3: Polynomial fit -} This method directly performs a
polynomial fit to the observed sequences in the CMD. The degree of the
polynomial has been chosen as a compromise between having an
adequate ability to reproduce the shape of the MS in the CMD and the
need of limiting the number of coefficients. We thus produced 11
different MRLs per CMD by varying the degree of the polynomial from 5
to 15, in step of 1.

The visual inspection of the MRLs derived with the three methods for
each cluster in each color combination shows that they are all similar
and they provide equivalently good representations of the cluster MS.
However, in order to test the effect of adopting slightly different
MRLs we determined the MS-TO and the MS-saddle in all the obtained
MRLs, separately. In doing this we adopted the following
prescriptions: {\it i)} The MS-TO has been defined as the bluest point of the
MS. This is one of the classical and most used definitions and it
allows an easy estimate of the error on its determination. 

{\it ii)} The MS-saddle
 is defined as the point of minimum curvature along the MS. To
measure it we used both an analytical and a geometric method. 

{\it Analytical method:} we defined the MS-saddle as the point where
the second derivative of the MRL is equal to zero. 

{\it Geometric method:} we adopted the ``circumference method'' described in
\citet{Mas16}. It consists in determining the circumference that
connects each point of the MRL with the two adjacent/contiguous points
located at $\pm$ 0.5 mag, and then adopt as MS-saddle the point
corresponding to the circumference with the largest radius (minimum
curvature). Following the suggestion in \citealp{Mas16}, we tested the
robustness of such a procedure by changing the distance in magnitude
among the three points on the MRL from 0.1 mag to 0.8 mag, in steps of
0.1 mag. This provided us with 8 different estimates of the MS-saddle
point for each MRL. In all cases the 8 measures nicely agree
(typically within $\pm 0.03$ mag). Hence, their average value has been
adopted as final ``geometric'' measure of the MS-saddle.

\begin{figure}
\epsscale{1.}
\plotone{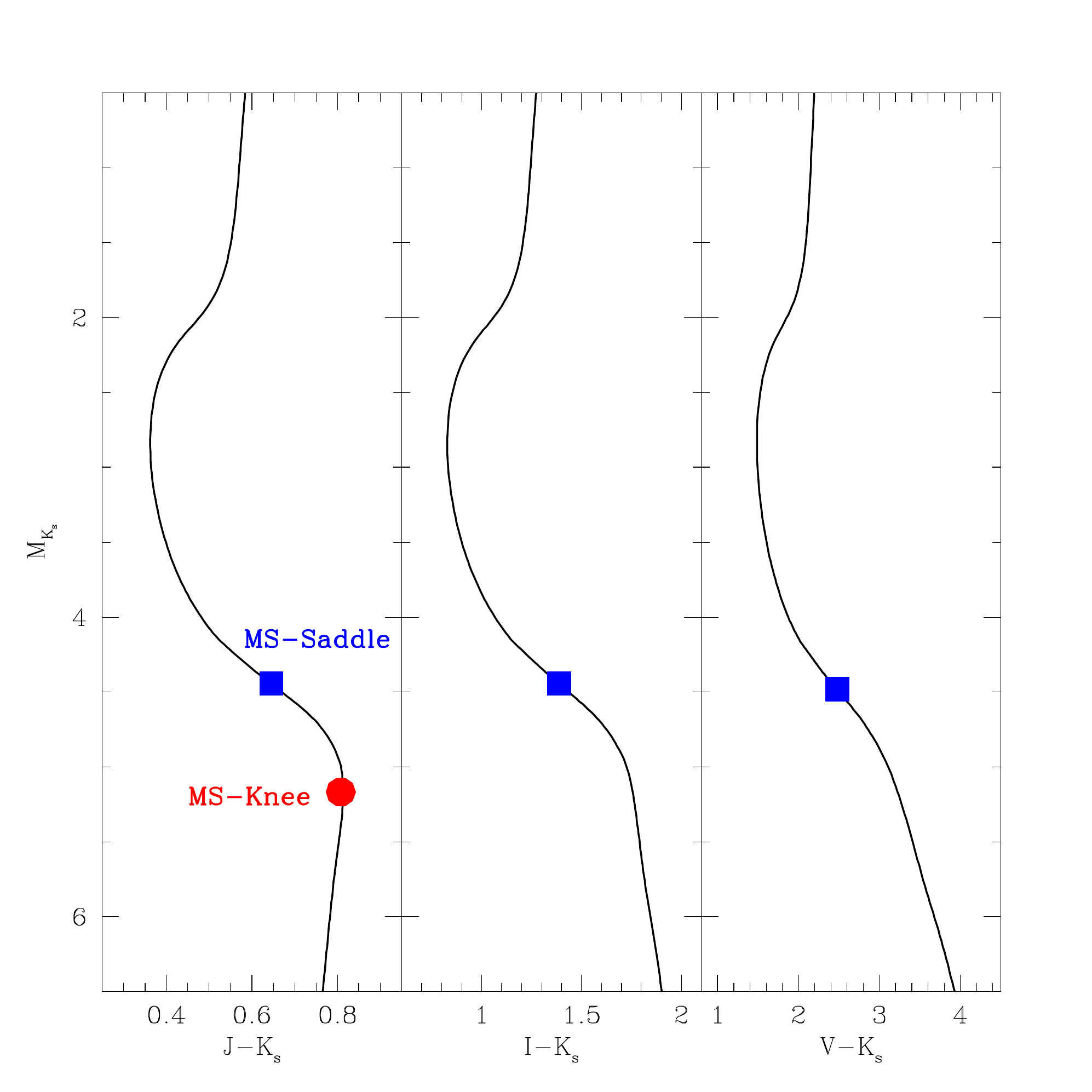}
\caption{Location of the MS-knee (red circle) and MS-saddle (blue
  square) marked along the same isochrone plotted in Figure \ref{fig2}, but here
  shown in CMDs with three different colors: from left to
  right, ($K_{s}, J-K_{s}$), ($K_{s}, I-K_{s}$), and ($K_{s}, V-K_{s}$). The MS-knee only occurs
   in the pure NIR-CMD. The MS-saddle,
  instead, can be defined in all the considered diagrams.}
\label{fig3}
\end{figure}

\begin{figure}
\epsscale{1.}
\plotone{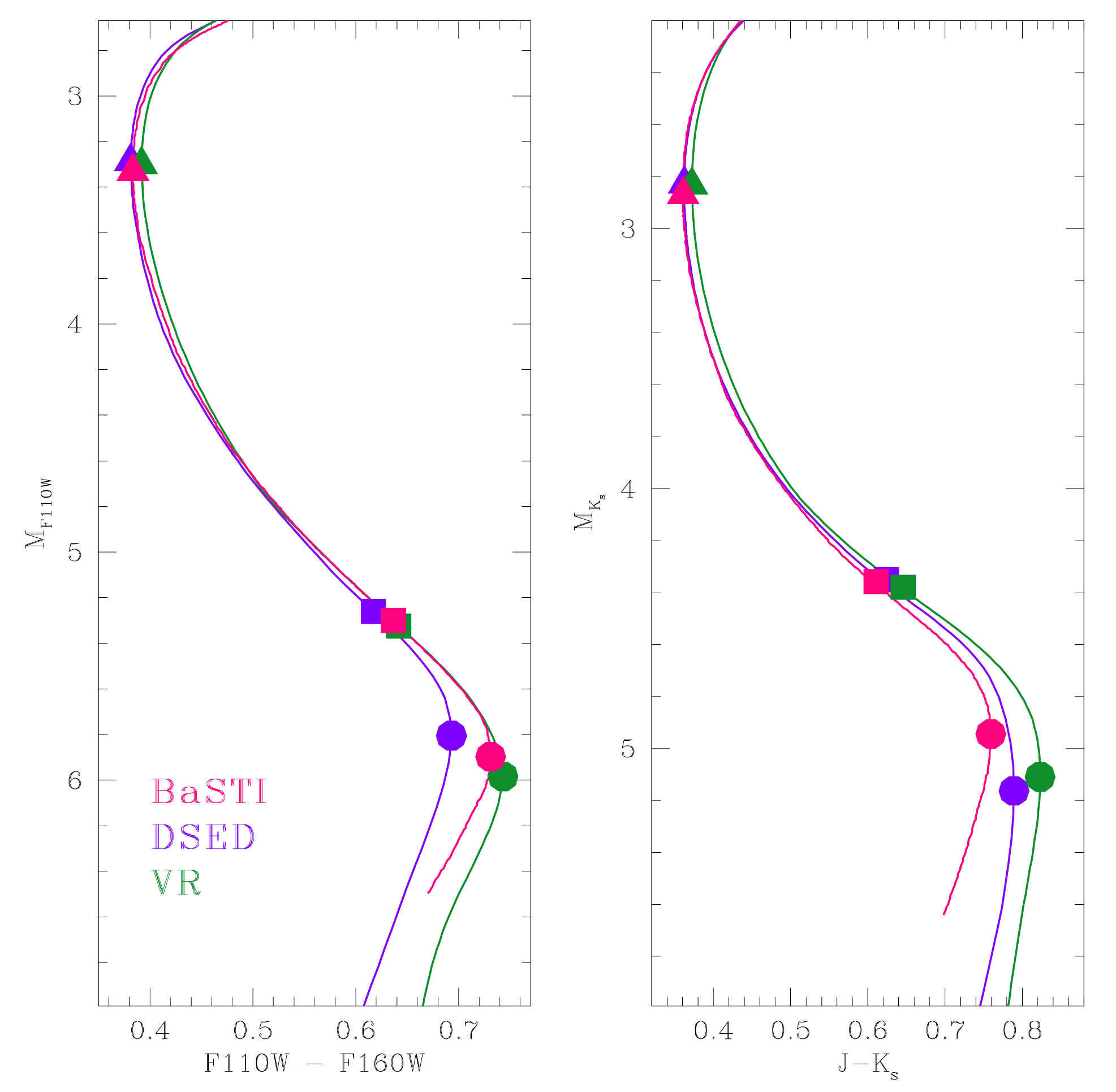}
\caption{Location of the MS-TO (triangles), MS-knee (circles) and
  MS-saddle (squares) marked along 12 Gyr old isochrones from three
  different sets of theoretical models: BaSTI (magenta), DSED (violet) and
  VR (green) in the ($K_{s}, J-K_{s}$, {\it left panel}) and ($F110W, F110W-F160W$, {\it right panel}) filter combinations. The three models 
  predict different locations of the MS-knee, while they agree on the 
  color-magnitude position of the MS-saddle. This clearly illustrates 
  that the two features are different.}
\label{fig4}
\end{figure}

\begin{figure}
\epsscale{1.}
\plotone{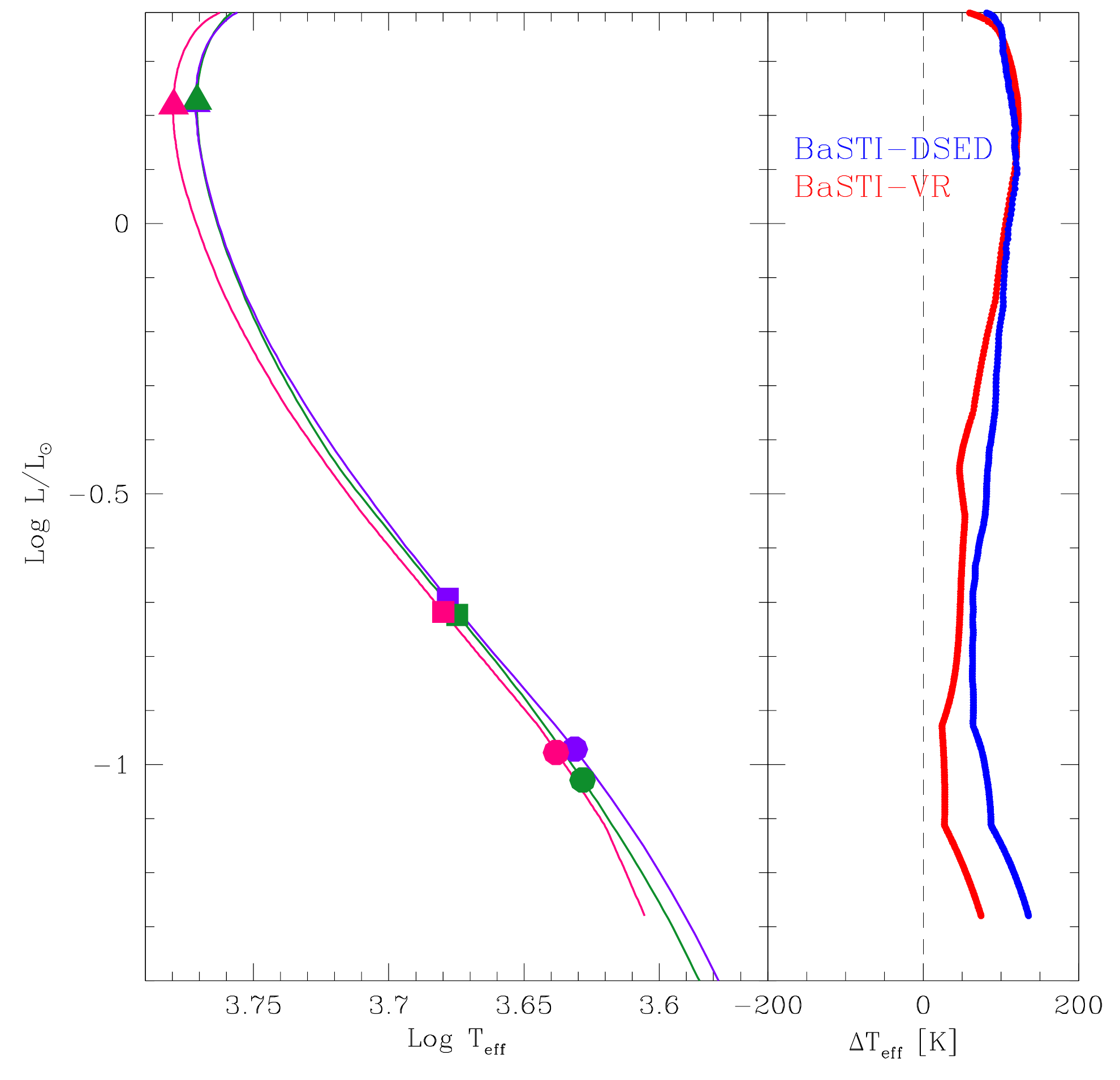}
\caption{{\it Left panel:} Color and magnitude of the MS-saddle (squares) and MS-knee (circles) points shown in Figure 4 are translated here in temperature and luminosity, respectively. {\it Right panel:} At fixed luminosity, the $\Delta T_{eff}$ between two pairs of models (BaSTI-DSED) and (BaSTI-VR) is presented in blue and red, respectively.}
\label{fig5}
\end{figure}

\begin{figure}
\epsscale{1.}
\plotone{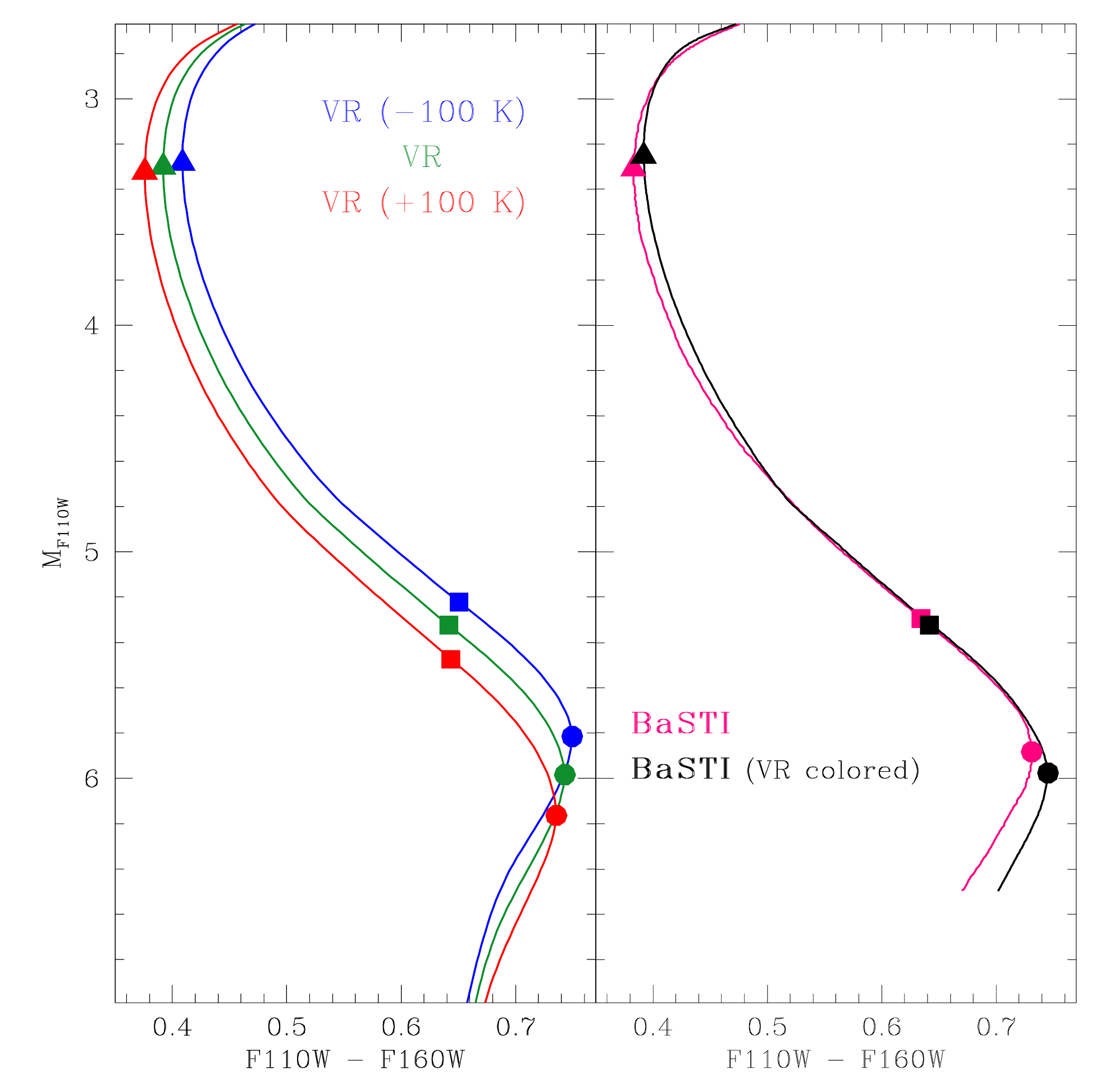}
\caption{{\it Left panel:} A VR isochrone is adjusted to hotter (red line) and cooler (blue line) temperatures by 100 K with respect to the normal one (green line) and then transformed to the observed ($F110W, F110W-F160W$) plane. {\it Right panel:} In the same filter combination, a BaSTI isochrone colored using its usual BCs is shown in magenta, compared to the same isochrone colored using the \citet{CV14} BCs (in black). Triangles, squared and circles indicate the MS-TO, MS-saddle and MS-knee, respectively, in both panels.} 
\label{fig6}
\end{figure}

\begin{figure}
\epsscale{1.}
\plotone{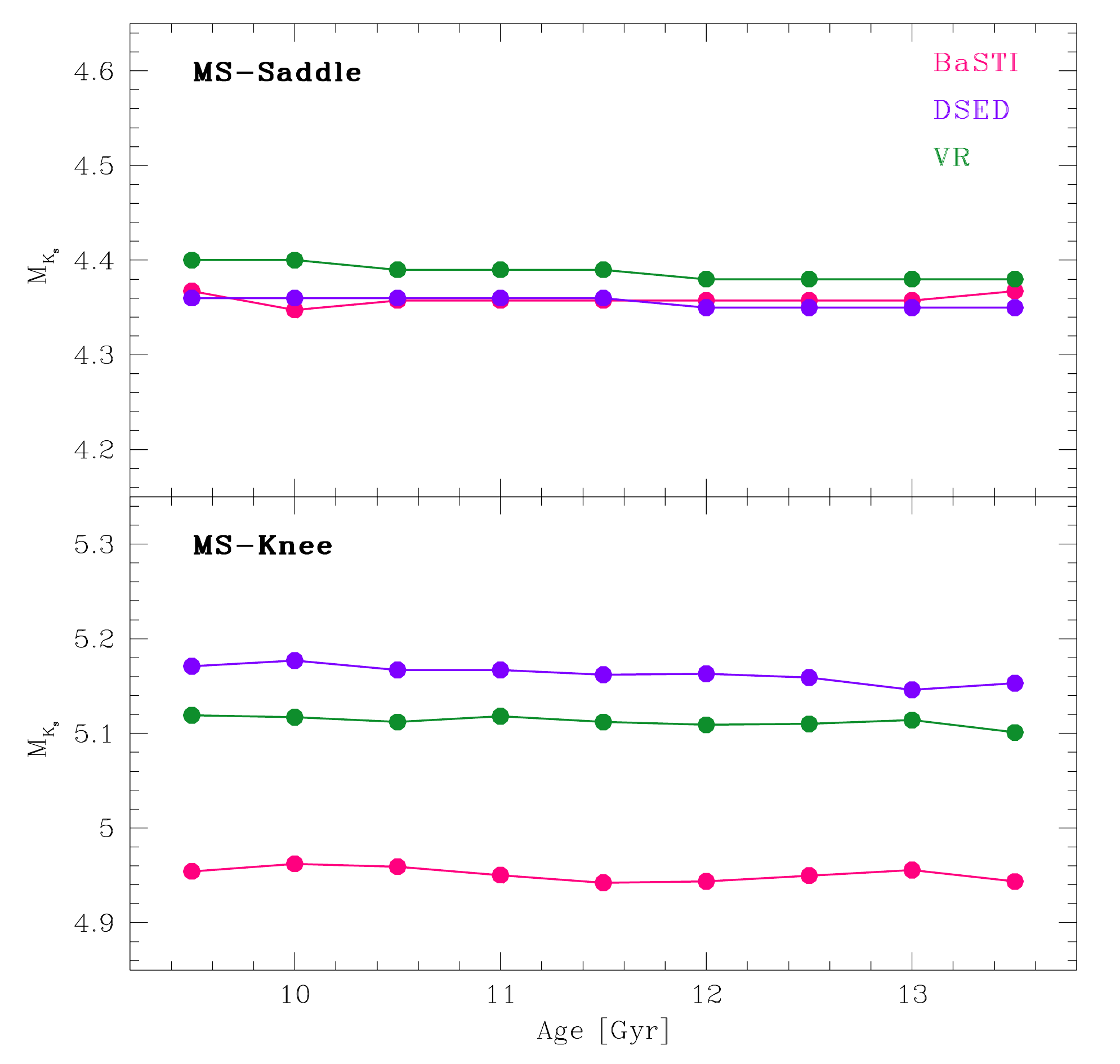}
\caption{{\it Upper panel--} Dependence of the MS-saddle $K_{s}$-band
  magnitude on cluster age for the three different sets of adopted
  isochrones (BaSTI, DSED and VR in magenta, violet and green,
  respectively). Ages vary from 9.5 to 13.5 Gyr in steps of 0.5
  Gyr. At fixed chemical composition, the MS-saddle $K_s$-band magnitude
  is independent of age and shows small dependence on
  the adopted model.  {\it Lower panel --} The same as in the
  upper panel, but for the MS-knee. In this case, different models
  predict significantly different values of the $K_s$-band magnitude of the
  MS-saddle (see Figure \ref{fig4} for further details).}
\label{fig7}
\end{figure}

\begin{figure}
\epsscale{1.}
\plotone{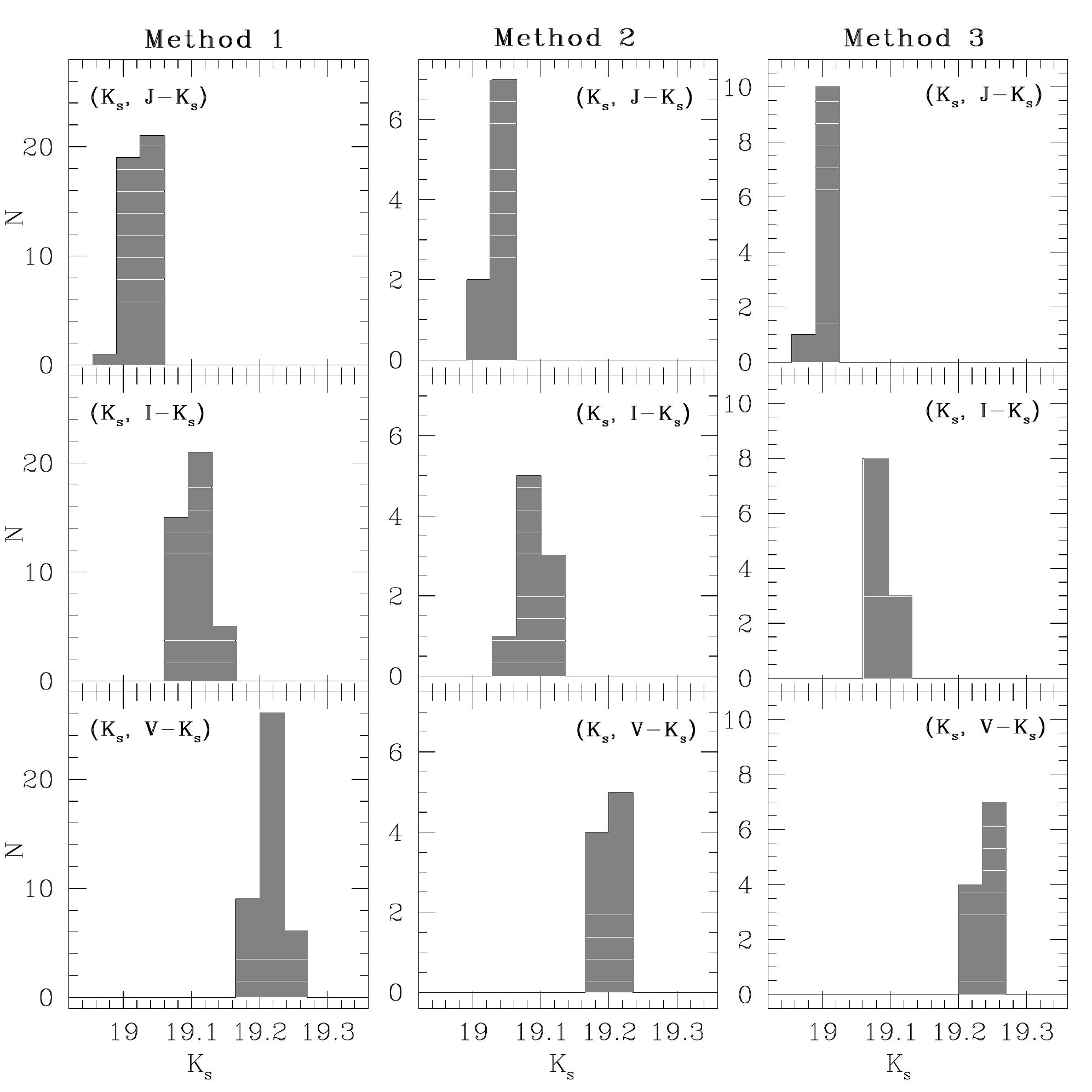}
\caption{Histograms of the MS-saddle $K$-band magnitude measured in
  the ($K_{s}, J-K_{s}$), ($K_{s}, I-K_{s}$), and ($K_{s}, V-K_{s}$) CMDs (upper, middle, and
  bottom rows, respectively), for NGC 6624. Multiple measures of this value have
  been obtained in each case because the MS MRL has been determined
  with different methods (see Sect. 3.1): Method 1 - static bins
  (left column, providing 41 MRLs and 41 measures of the MS-saddle
  magnitude), Method 2 - dynamic bins (central column, 9 measures),
  and Method 3 - Polynomial fit (right column, 11 measures).}
\label{fig8}
\end{figure}

\begin{figure}
\epsscale{1.}
\plotone{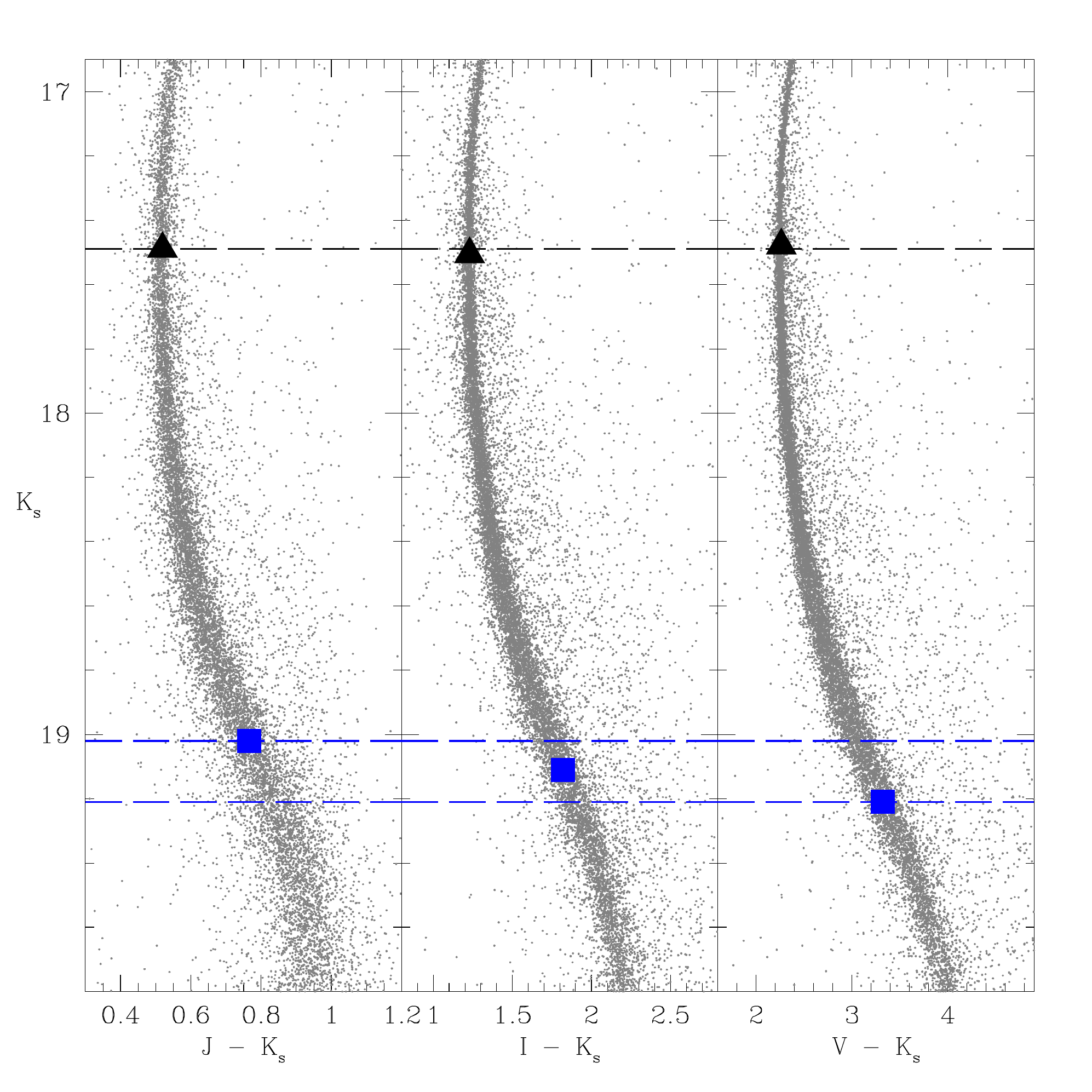}
\caption{Location of the MS-TO (black triangle) and MS-saddle (blue
  square) in the three CMDs available for NGC 6624. The horizontal
  black dashed line marks the MS-TO level, the horizontal dashed blue
  lines the two extreme values of the MS-saddle, which differ by 0.2 mag.}
\label{fig9}
\end{figure}

The application of different methods for determining the cluster MRL
and the use of different CMDs affect the estimate
of the MS-saddle magnitude as illustrated in Figure \ref{fig8} for NGC
6624. The results are shown for the ``geometric'' measure of the 
MS-saddle; those obtained from the analytical method are fully
consistent.  According to the description above, we have derived 41, 9
and 11 MLRs from methods 1, 2 and 3, respectively, for each color.
Hence, the three panels in the upper row of Figure \ref{fig8} show the
distribution of the 41, 9 and 11 values of the MS-saddle $K_{s}$-band
magnitude determined in the ($K_{s}, J-K_{s}$) CMD from the three methods. The
middle and bottom rows show the analogous results obtained from the
($K_{s}, I-K_{s}$) and ($K_{s}, V-K_{s}$) diagrams. For a given CMD (along
each row in the figure) the estimated magnitude of the MS-saddle is
essentially independent of the method adopted to determine the MRL (all the
measures agree within $\pm 0.04$ mag). However, for any fixed method
to determine the MRL, the magnitude of the MS-saddle varies by
0.2-0.25 mag when different CMDs are considered (i.e., along each
column in Figure \ref{fig8}). In particular, the MS-saddle $K_{s}$-band magnitude shows a
systematic trend with the color, becoming increasingly
fainter for color combinations that involve filters at shorter
wavelengths. This is also illustrated in Figure \ref{fig9}, where the MS-saddle
point (determined as the average of the 41 measures resulting from the
``static bins'' and the ``geometric'' methods) is marked with a
blue square in each of the three available CMDs. This trend might be
an indirect effect induced by the presence and the
relevance of the MS-knee on the shape of the MS MRL. In fact the
presence of a clear knee in the ($K_{s}, J-K_{s}$) CMD (see the left
panel of Figure \ref{fig3}) might require an ``early change'' in the curvature of
the MRL (i.e. the inflection point must occur at relatively bright
luminosity). Instead, as soon as the knee disappears becoming a
``light bend'' in the MRL (middle and right panels of Figure \ref{fig3}), the
change in the MRL curvature becomes ``less pronounced'' (i.e. the
inflection point tends to slide to fainter
luminosities). At odds with the systematic drift
of the MS-saddle magnitude with the color, the measures of
the MS-TO stay nicely stable (within 0.02 mag) independently of the
considered CMD. This is also apparent in Figure \ref{fig9} (black triangles).
If the CMD location of the MS-saddle significantly depends on the details of the
MS MRL morphology, theoretical isochrones have to predict such a trend accordingly. 
If not, this might introduce some systematic in estimating age using this feature.

Since the magnitude of the MS-saddle does not depend on the method used
to determine the MRL, in the following we will consider only the
values obtained from Method 1 ({\it static bins}), which also provides
the largest statistics (41 data points). The magnitudes of the MS-TO
and MS-saddle (and their uncertainties) obtained as the average (and
the standard deviation) of the 41 measurements in each of the
available color combinations are listed in Table 2. For 47
Tucanae the listed magnitudes are in the $F110W$ band, while for
NGC 6624 they are in the $K_{s}$ band.
The last two columns of the table list the values obtained from the
two methods (analytic and geometric) used to identify the MS-saddle
point along the MRL. As can be seen, the results are fully consistent
and in the following we will thus adopt the values obtained from the
{\it geometric approach}, which has been already used in the
literature.

We finally estimated the parameter $\Delta^{\rm saddle}_{\rm TO}$ from
the difference between the adopted MS-saddle and MS-TO magnitudes and 
we derived the uncertainties by taking into account: {\it (1)} the error associated to the MS-TO
determination; {\it (2)} the uncertainty related to the MS-saddle
point and {\it (3)} the photometric error affecting both
parameters.
Photometric errors are an additional source of
uncertainty in the derivation of the MS-saddle and the MS-TO positions,
so they have to be taken into account. They have been computed as the
average of the photometric errors (in $F110W$ and in $K_s$ for 47 Tucanae
and NGC 6624, respectively) of all the stars falling within $\pm$ 0.1
magnitudes from the MS-TO and the MS-saddle points. In the case of 47
Tucanae, we obtained $\sigma_{F110W, \rm{MSTO}} = 0.002$ mag and
$\sigma_{F110W, \rm{saddle}} = 0.005$ mag. For NGC 6624 we found $\sigma_{K_s,
\rm{MSTO}} = 0.005$ mag and $\sigma_{K_s, \rm{saddle}} = 0.035$ mag for
all the color combinations. Photometric errors do not actually have
any major impact on the final uncertainties of $\Delta^{\rm saddle}_{\rm TO}$, because
they are relatively small with respect to other uncertainties. Moreover, MRLs are intrinsically 
affected by their contribution because the broadening
of the MS is mainly function of photometric errors.
The final values and uncertainties of
$\Delta^{\rm saddle}_{\rm TO}$ are listed in the last column of Table 2. As
can be seen, this parameter can change by $0.2$ mag, because of the
sensitivity of the MS-saddle point to the selected color.

\section{Absolute GC ages derived from the MS-saddle}
In this section we estimate the absolute age of the two studied clusters from the
 $\Delta^{\rm saddle}_{\rm TO}$ parameter. In order to provide a set of analytical relations linking the cluster
age and the $\Delta^{\rm saddle}_{\rm TO}$ parameter, we considered
models covering a meaningful range of ages and metallicities. In
particular, the cluster age was sampled in steps of 0.5 Gyr, from 9.5
Gyr to 13.5 Gyr (this is a reasonable age range for both clusters,
according to previous literature estimates; see \citealp{Cor16} for 47
Tucanae, and \citealp{Sara16} and references therein for NGC 6624).

Concerning metallicity, given that for both clusters 
slightly different values have been reported in the literature
(see, e.g., \citealp{Car09, Car10, Val04a, Val04b, Val11}),
we considered models spanning a range of $\pm 0.1$ dex in steps of 0.05 dex
around the quoted values of [Fe/H] for 47 Tucanae and NGC 6624 (see Table 1).
Thus a grid of isochrones with the adopted ages and metallicities has
been built. 

To estimate the theoretical values of $\Delta^{\rm
saddle}_{\rm TO}$ in all the considered filter combinations, we re-sampled the
BaSTI, DSED and VR models adopting the same magnitude steps used for
the determination of the MRLs with the ``static bins'' method.  Then,
the magnitudes of the MS-TO and the MS-saddle points have been derived
by adopting the same geometric approach used for the observed CMDs.
 
At the end of the procedure we thus have a grid of points
corresponding to one value of $\Delta^{\rm saddle}_{\rm TO}$ for each
of the considered isochrones of different ages and metallicities. To
determine the absolute cluster age (in Gyr) as a function of 
$\Delta^{\rm saddle}_{\rm TO}$ and metallicity we use a linear bi-parametric
fit. The resulting analytic relations are listed in Table 3 for the
three adopted families of stellar models and for each filter 
combination available in our observational datasets. 

The BaSTI isochrones appear to be the most sensitive to
the parameter $\Delta^{\rm saddle}_{\rm TO}$: an uncertainty of $\pm$
0.1 dex in $\Delta^{\rm saddle}_{\rm TO}$ produces uncertainties up to
2 Gyr in age. 

The DSED isochrones appear to be the most sensitive to the metallicity: an uncertainty of $\pm$ 0.1 dex in
metallicity produces an uncertainty of $\pm$ 0.8-0.9 Gyr in age. 

It is also worth noticing that by enlarging the baseline color towards the blue, 
younger ages are obtained. The reason has to be find in the $K_{s}$ magnitude variation of the MS-saddle as 
a function of the color baseline, already discussed in Section 4. Stellar models miss to predict such a trend 
with the actual consequence of making GCs gradually younger, using the $\Delta^{\rm saddle}_{\rm TO}$ parameter. 
For instance, the VR isochrones are the most sensitive to the adopted color and 
by moving from ($J-K_s$) to ($V-K_s$), up to 2.3 Gyr younger ages are derived.

\begin{figure}
\epsscale{1.}
\plotone{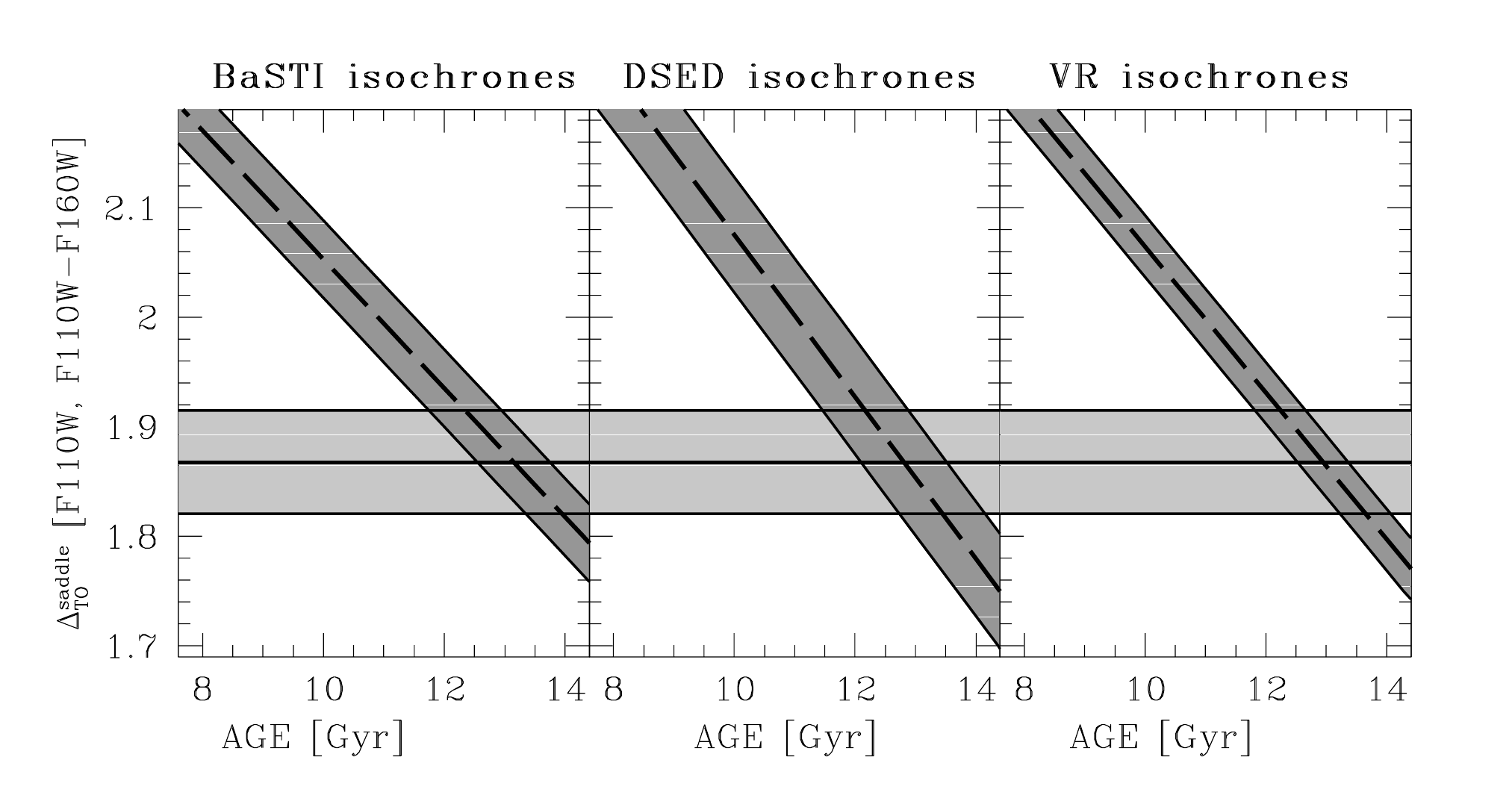}
\caption{Predicted relations between age (in Gyr) and the parameter
  $\Delta ^{\rm saddle}_{\rm TO}$ obtained from BaSTI, DSED and VR
  isochrones. The dashed
  lines are the theoretical relations computed at the chemical composition of
  47 Tucanae (\citet{Cor16} and references therein); the dark grey
  regions surrounding each dashed line mark the variation induced by
  changes of $\pm 0.1$ dex in the adopted metallicity. The solid line
  and grey region in each panel mark the observed value and
  uncertainty of the $\Delta ^{\rm saddle}_{\rm TO}$ parameter
  measured in the ($F110W, F110W-F160W$) CMD of 47 Tucanae.}
\label{fig10}
\end{figure}

\begin{figure}
\epsscale{1.}
\plotone{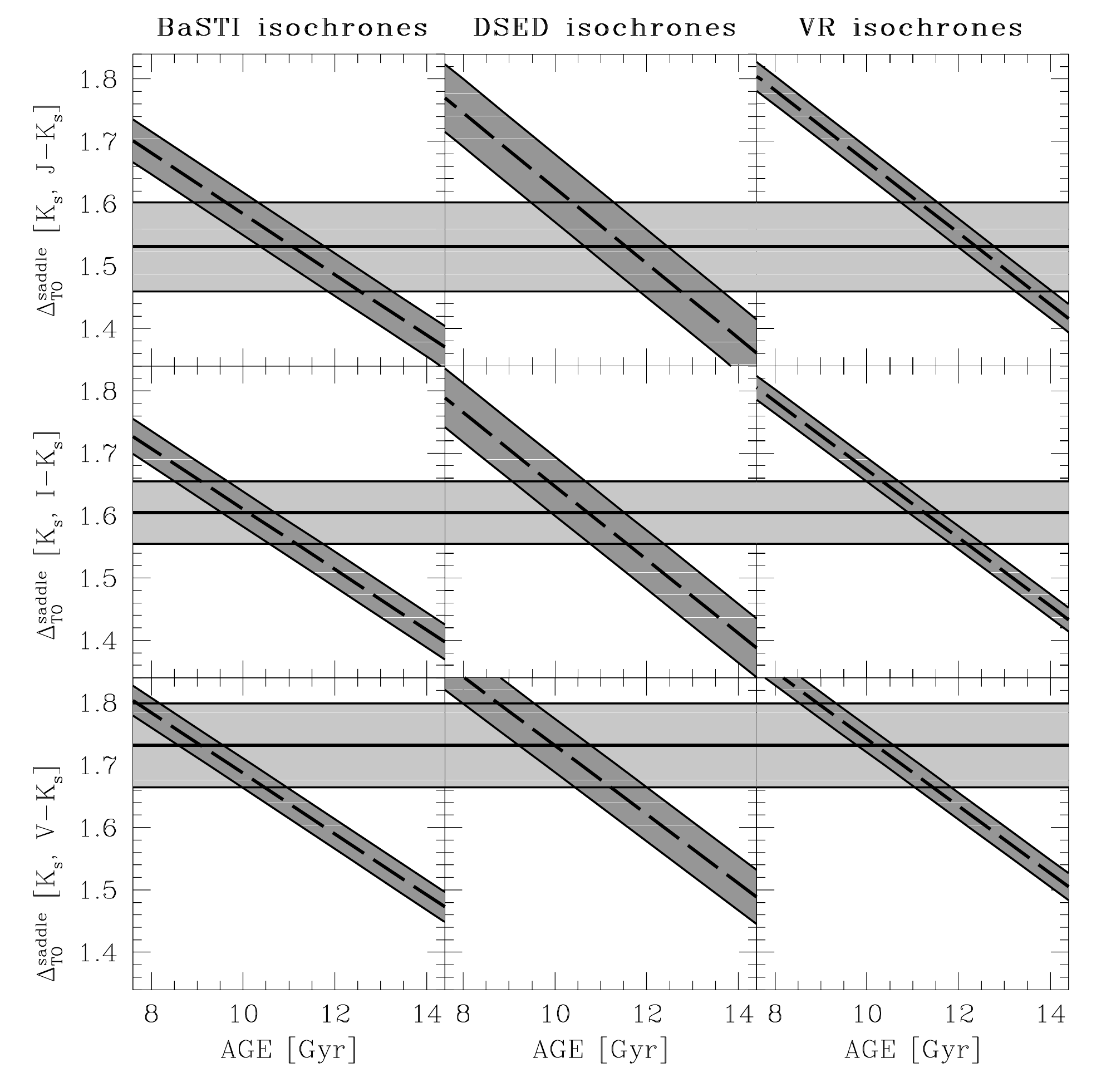}
\caption{The same as in Figure \ref{fig10} but 
in the ($K_{s}, J-K_{s}$), ($K_{s}, I-K_{s}$), and ($K_{s}, V-K_{s}$) for NGC 6624. In this case the
  theoretical relations and the observed values have been determined
  in the three available CMDs (see labels).} 
\label{fig11}
\end{figure}

The results of this analysis are also shown in Figure \ref{fig10}
for 47 Tucanae and Figure \ref{fig11} for NGC 6624. 
Each panel refers to one model and one color combination.
The dashed lines 
correspond to the analytical relations at the nominal cluster metallicity, and the 
surrounding dark grey regions encompass a  $\pm$ 0.1 dex variation 
in metallicity.

In Figures \ref{fig10} and \ref{fig11} we also plot the observed values of $\Delta^{\rm saddle}_{\rm TO}$ (horizontal solid lines) 
and their uncertainty (light grey region). 
From the intersection of
the observed values with the theoretical expectations we could finally derive the
absolute age of the clusters and their uncertainties. The
results are listed in Table 4. Errors on the age have been computed by using the uncertainties in the measured 
$\Delta^{\rm saddle}_{\rm TO}$ quoted in Table 2 and of $\pm$ 0.1 dex in metallicity.

For 47 Tucanae we obtain age values between 12.8 and 13.1 Gyr with an uncertainty $\le$ 1.5 Gyr.
These age values are larger than the 11.6 $\pm$ 0.7 Gyr recently obtained by \citet{Cor16}, 
although still marginally consistent within our errors. 
For NGC 6624 we obtain age values between 9.9 and 12.5 Gyr and $\le 2.1$ Gyr uncertainty, 
depending on the adopted color and model. As a comparison, the recent determination of \citet{Sara16} for this cluster is of about 12.0 $\pm$ 0.5 Gyr.

If less conservative assumptions for the $\Delta^{\rm saddle}_{\rm TO}$ and metallicity uncertainties are made
as done in previous works \citep[see e.g.][]{DiC15, Mas16}, 
for a selected model and CMD, errors can decrease to $\approx$1~Gyr or even below.

\section{Conclusions}
In this paper we presented a detailed analysis of 
the MS-knee and the MS-saddle features for absolute GC
age determinations.
To this end we used three different families of stellar models and
deep NIR observations of the two metal-rich Galactic GCs 47 Tucanae and NGC 6624.
For NGC 6624 deep optical photometry was also available, thus allowing to 
explore the behavior of the two features in different colors. 
The main conclusions of the paper can be summarized as follows: 
\begin{itemize}
\item[1.] {\it A more-suitable definition of the MS-knee -} Because in NIR
CMDs the low-mass end of the MS bends to the blue and forms a ``knee''
(see Figure \ref{fig2}), this feature needs to be defined consistently as the
reddest point along the MS. Unfortunately, however, this definition
strictly holds only for pure NIR CMDs (see Figure \ref{fig3}). Moreover, the
predicted magnitude and color of the MS-knee are still significantly
model-dependent (see Figures \ref{fig4}, \ref{fig5} and \ref{fig6}), thus preventing a firm absolute age
determination based on this feature. 
Hence, 
a theoretical effort is required 
to remove the discrepancies among different isochrones in the location of the MS-knee
(which can lead to several Gyr differences in age), by identifying the input
physics and/or the $T_{eff}$ - color transformations responsible for them. 

\item[2.] {\it A MS-saddle, not a knee -} 
The MS-knee was
originally defined as the point of minimum curvature along the MS MRL
\citep{Bon10}. Figure \ref{fig2} shows that the point where the MS bends
to the blue (the so-called ``knee'') does not coincide with the {\it
inflection point}, where the MS MRL has its minimum curvature
because it changes its curvature from convex to concave. Hence, here
we more appropriately named this latter point {\it ``MS-saddle''}. The
MS-saddle is 0.7 mag brighter than the MS-knee and typically samples a
mass $\sim$0.1 $M_{\odot}$ higher with respect to the MS-knee. 
Figure \ref{fig4} demonstrates that the MS-saddle is insensitive to
the morphology and the location of the MS-knee, since
different bends of the MS (hence different magnitudes for the 
MS-knee) correspond to very similar MS-saddle points, with similar   
luminosities and colors. Hence the MS-saddle cannot be considered a
proxy of the MS-knee.
  
\item[3.] {\it The MS-saddle: a fragile feature -} At odds with the MS-knee, 
which has a physical nature, the MS-saddle is just a geometric point 
indicating a change in the curvature of the MS MRL. We performed
a detailed analysis of its properties by making use of deep NIR and
photometry of the GCs 47 Tucanae and NGC 6624. 
All the measures of the MS-saddle
magnitude obtained from the different methods turned out to agree
within $\pm 0.04$ mag, thus having a modest impact (at a sub Gyr-level) on the final age estimate. 

The analysis of NGC 6624 also offered the possibility to study the
location of the MS-saddle in CMDs with different colors,
namely the ($K_{s}, J-K_{s}$), ($K_{s}, I-K_{s}$) and ($K_{s}, V-K_{s}$) diagrams. We found
that the $K_{s}$-band magnitude of the MS-saddle changes by 0.2-0.25 mag
when different colors are considered (see Figure \ref{fig8}). Moreover,
a systematic trend has been detected, with the MS-saddle becoming
brighter with color baseline extending towards bluer filters
(see Figures \ref{fig9} and \ref{fig11}, and Table 4), while in the same CMDs the absolute magnitude of the MS-TO stays
nicely constant (within 0.02 mag).
Such a color dependence of the MS-saddle location is not predicted by theoretical isochrones,
thus making it an unreliable anchor to estimate absolute ages.

\item[4.] {\it The MS-saddle: not an improvement -} 
State-of-the-art absolute age determination of GCs using MS-fitting methods \citep[see e.g.][which use the morphology of the sequence to constrain also reddening, distance and metallicity]{Cor16} 
already provide values with sub-Gyr uncertainties. 
The age values derived from the $\Delta_{\rm TO}^{\rm saddle}$ parameter can have similar sub-Gyr uncertainties in the most
optimistic assumption of a few hundredths mag uncertainty in the positioning of the MS-saddle and a few hundredths dex 
uncertainty for the cluster metallicity. 
Moreover, the systematic dependence of the inferred ages with the selected colors 
(i.e. younger ages for color baselines more extended to the blue), makes impossible 
to set a unequivocal absolute age scale.
\end{itemize}

The next generation of space and ground-based telescopes, like the
James Webb Space Telescope (JWST) and the European Extremely Large
Telescope (ELT), is expected to give a significant impulse to NIR
observations. In particular, the low-mass MS could be studied in great
detail in many more GCs. This will allow a more precise observational
characterization of the MS-knee, to better constrain the structure of
extreme low-mass stars and the input physics for their modeling, in order 
to finally solve the current discrepancies among stellar models. 
A major improvement in terms of absolute GC ages is also expected from the significant
refinement of the distance determinations from the Gaia mission \citep{Gaia16}.
This perspective definitively suggests that the absolute GC age dating methods
will live very soon a renewed youth.

\section{Acknowledgements}
We thank the anonymous referee for the careful reading of the paper and the useful suggestions that greatly helped to better present our results.
The authors warmly thank Giuseppe Bono for useful comments and discussions. 
SV gratefully acknowledges the support provided by Fondecyt Regular project n. 1170518. CMB gratefully acknowledges the support provided Fondecyt Regular project n. 1150060.
\clearpage

\begin{table}[!ht]
	\begin{center}
		\caption{The abundance of some key elements for the BaSTI, DSED and VR models adopted in this work, according to their solar mixtures.}
		\label{tab4}
		\footnotesize
		\begin{tabular}{ccccc}
			\\
			\hline
			\hline
			cluster &  & BaSTI isochrone & DSED isochrone & VR isochrone\\
			\hline
			47 Tucanae & [Fe/H] & -0.70 & -0.64 & -0.65 \\
			... & log N(O) & 8.57 & 8.59 & 8.44 \\
			... & log N(Mg) & 7.28 & 7.34 & 7.35 \\
			... & log N(Si) & 7.25 & 7.31 & 7.26 \\
			... & Z & 0.0080 & 0.0075 & 0.0060 \\
			... & Y & 0.2560 & 0.2571 & 0.2571 \\
			... & [M/H] & -0.353 & -0.355 & -0.350 \\			
			\hline
			\hline
			NGC 6624 & [Fe/H] & -0.60 & -0.54 & -0.55 \\
			... & log N(O) & 8.67 & 8.69 & 8.54 \\
			... & log N(Mg) & 7.38 & 7.44 & 7.45 \\
			... & log N(Si) & 7.35 & 7.41 & 7.36 \\
			... & Z & 0.0100 & 0.0095 & 0.0075 \\
			... & Y & 0.2590 & 0.2604 & 0.2600 \\
			... & [M/H] & -0.253 & -0.250 & -0.250 \\
			\hline
		\end{tabular}
	\end{center}
	Notes: Solar mixtures from \citet{grev93}, \citet{grev98} and \citet{Asp09} for BaSTI, DSED and VR models, respectively. [M/H] values are computed as $(Z/X)_{iso} - (Z/X)_{\odot}$, where $(Z/X)_{\odot}$=0.0245 for BaSTI, 0.0231 for DSED and 0.0181 for VR isochrones. 
\end{table}

\begin{table}[!ht]
\begin{center}
\caption{MS-TO and MS-saddle magnitudes, and their difference
  $\Delta^{\rm saddle}_{\rm TO}$ in the $F110W$ (for 47 Tucanae) and in the
  $K_{s}$ band (for NGC 6624). The listed values are the average of the 41
  measures determined by adopting the {\it static bin} and the {\it
    geometric approach} (see Sect. 3.1).}
\label{tab1}
\footnotesize
\begin{tabular}{cccccc}
\\
\hline
\hline
cluster & CMD & MS-TO & MS-saddle  & MS-saddle & $\Delta^{\rm saddle}_{\rm TO}$\\
        &     &      & (analytic) & (geometric) & \\
\hline
47 Tucanae & ($F110W, F110W-F160W$) & 16.64 $\pm$ 0.03 & 18.53 $\pm$ 0.03 & 18.51 $\pm$ 0.04 & 1.87 $\pm$ 0.05 \\
\hline
NGC 6624 & ($K_{s}, J-K_{s}$) & 17.49 $\pm$ 0.03 & 19.08 $\pm$ 0.05 & 19.02 $\pm$ 0.05 & 1.53 $\pm$ 0.07 \\
... & ($K_{s}, I-K_{s}$) & 17.51 $\pm$ 0.03 & 19.15 $\pm$ 0.05 & 19.11 $\pm$ 0.04 & 1.61 $\pm$ 0.05 \\
... & ($K_{s}, V-K_{s}$) & 17.48 $\pm$ 0.02 & 19.22 $\pm$ 0.03 & 19.21 $\pm$ 0.03 & 1.73 $\pm$ 0.06 \\
\hline
\end{tabular}
\end{center}
\end{table}

\begin{table}[!ht]
\begin{center}
  \caption{Analytic relations between age ($t$, in Gyr), $\Delta^{\rm
      saddle}_{\rm TO}$ and metallicity derived from the three adopted
    families of stellar models and the color combinations available in
    our observational datasets of 47 Tucanae and NGC 6624.}
\label{tab2}
\footnotesize
\begin{tabular}{ll}
\\
\hline
\hline   
 &   \,\,\,\,\,\,\,\,\,\,\,\,\,\,\,\,\,\,\,\,\,\,\,\,\,\,\,\,\,\,\,\,\,\,  BaSTI models \\
 ($F110W, F110W-F160W$) & $t$ [Gyr] $= 40.66 (\pm 0.82) - 16.99 (\pm 0.46) \times \Delta^{\rm saddle}_{\rm TO} - 6.02 (\pm 0.41) \times [Fe/H]$ \\
\,\,\,\,\,\,\,\,\,\,\,\,\,\,\,\,\,\,\,\, ($K_{s}, J-K_{s}$) & $t$ [Gyr] $= 38.34 (\pm 0.63) - 20.56 (\pm 0.47) \times \Delta^{\rm saddle}_{\rm TO} - 7.05 (\pm 0.36) \times [Fe/H] $     \\
\,\,\,\,\,\,\,\,\,\,\,\,\,\,\,\,\,\,\,\, ($K_{s}, I-K_{s}$) & $t$ [Gyr] $= 39.74 (\pm 0.68) - 20.63 (\pm 0.48) \times \Delta^{\rm saddle}_{\rm TO} - 5.81 (\pm 0.36) \times [Fe/H] $     \\
\,\,\,\,\,\,\,\,\,\,\,\,\,\,\,\,\,\,\,\, ($K_{s}, V-K_{s}$) & $t$ [Gyr] $= 41.70 (\pm 0.34) - 20.54 (\pm 0.23) \times \Delta^{\rm saddle}_{\rm TO} - 4.92 (\pm 0.17) \times [Fe/H] $     \\
 \hline
 &  \,\,\,\,\,\,\,\,\,\,\,\,\,\,\,\,\,\,\,\,\,\,\,\,\,\,\,\,\,\,\,\,\,\,\  DSED models \\
 ($F110W, F110W-F160W$) & $t$ [Gyr] $= 33.51 (\pm 0.68) - 14.94 (\pm 0.45) \times \Delta^{\rm saddle}_{\rm TO} - 7.11 (\pm 0.44) \times [Fe/H]$ \\
\,\,\,\,\,\,\,\,\,\,\,\,\,\,\,\,\,\,\,\, ($K_{s}, J-K_{s}$) & $t$ [Gyr] $= 32.13 (\pm 0.87) - 16.62 (\pm 0.65) \times \Delta^{\rm saddle}_{\rm TO} - 9.03 (\pm 0.65) \times [Fe/H] $ \\
\,\,\,\,\,\,\,\,\,\,\,\,\,\,\,\,\,\,\,\, ($K_{s}, I-K_{s}$) & $t$ [Gyr] $= 33.60 (\pm 0.59) - 16.95 (\pm 0.42) \times \Delta^{\rm saddle}_{\rm TO} - 8.01 (\pm 0.40) \times [Fe/H]$ \\
\,\,\,\,\,\,\,\,\,\,\,\,\,\,\,\,\,\,\,\, ($K_{s}, V-K_{s}$) & $t$ [Gyr] $= 37.16 (\pm 0.45) - 18.13 (\pm 0.30) \times \Delta^{\rm saddle}_{\rm TO} - 7.83 (\pm 0.27) \times [Fe/H] $\\
 \hline
 &  \,\,\,\,\,\,\,\,\,\,\,\,\,\,\,\,\,\,\,\,\,\,\,\,\,\,\,\,\,\,\,\,\,\,\,\,  VR models \\
 ($F110W, F110W-F160W$) & $t$ [Gyr] $= 38.11 (\pm 0.85) - 14.94 (\pm 0.45) \times \Delta^{\rm saddle}_{\rm TO} - 4.20 (\pm 0.44) \times[Fe/H]$  \\
\,\,\,\,\,\,\,\,\,\,\,\,\,\,\,\,\,\,\,\, ($K_{s}, J-K_{s}$) & $t$ [Gyr] $= 37.02 (\pm 0.72) - 17.53 (\pm 0.47) \times \Delta^{\rm saddle}_{\rm TO} - 4.02 (\pm 0.39) \times [Fe/H]$  \\
\,\,\,\,\,\,\,\,\,\,\,\,\,\,\,\,\,\,\,\, ($K_{s}, I-K_{s}$) & $t$ [Gyr] $= 38.70 (\pm 0.45)- 18.29 (\pm 0.29)  \times \Delta^{\rm saddle}_{\rm TO} - 3.48 (\pm 0.23) \times [Fe/H]$   \\
\,\,\,\,\,\,\,\,\,\,\,\,\,\,\,\,\,\,\,\, ($K_{s}, V-K_{s}$) & $t$ [Gyr] $= 40.17 (\pm 0.48) - 18.58 (\pm 0.30) \times \Delta^{\rm saddle}_{\rm TO} - 3.99 (\pm 0.24) \times 
[Fe/H]$  \\
\hline
\end{tabular}
\end{center}
\end{table}

\begin{table}[!ht]
\begin{center}
\caption{Absolute ages of 47 Tucanae and NGC 6624 estimated from the
  measured values of $\Delta^{\rm saddle}_{\rm TO}$ and for the three
  adopted families of stellar models (BaSTI, DSED, VR; see text).}
\label{tab3}
\footnotesize
\begin{tabular}{ccccc}
\\
\hline
\hline
cluster & CMD & $t_{\rm BaSTI}$  & $t_{\rm DSED}$ & $t_{\rm VR}$ \\
        &     & (Gyr)           & (Gyr)        & (Gyr)\\
\hline
47 Tucanae & ($F110W, F110W-F160W$) & 13.1 $\pm$ 1.4 & 12.8 $\pm$ 1.5 & 13.0 $\pm$ 1.2 \\
\hline
NGC 6624 & ($K_{s}, J-K_{s}$) & 11.1 $\pm$ 2.1 & 11.6 $\pm$ 2.1 & 12.5 $\pm$ 1.7 \\
...      & ($K_{s}, I-K_{s}$) & 10.1 $\pm$ 1.6 & 10.7 $\pm$ 1.7 & 11.3 $\pm$ 1.3 \\
...      & ($K_{s}, V-K_{s}$) & 9.9 $\pm$ 1.8 & 10.0 $\pm$ 2.0 & 10.2 $\pm$ 1.6 \\
\hline
\end{tabular}
\end{center}
\end{table}

\end{document}